\documentclass{vki}

\usepackage{graphicx}
\usepackage{tabls}
\usepackage{multirow}
\usepackage{hyperref}
\usepackage{xcolor}
\usepackage{amsfonts}
\usepackage{amsmath}

\begin{document}


\pagenumbering{arabic}
\title{Challenges and Opportunities for Machine Learning in Fluid Mechanics}
\author{M. A. Mendez, J. Dominique, M. Fiore, F. Pino, \\P. Sperotto, J. Van den Berghe}
\date{	von Karman Institute for Fluid Dynamics}
\maketitle

\begin{abstract}
Big data and machine learning are driving comprehensive economic and social transformations and are rapidly re-shaping the toolbox and the methodologies of applied scientists. Machine learning tools are designed to learn functions from data with little to no need of prior knowledge. As continuous developments in experimental and numerical methods improve our ability to collect high-quality data, machine learning tools become increasingly viable and promising also in disciplines rooted in physical principles. These notes explore how machine learning can be integrated and combined with more classic methods in fluid dynamics. After a brief review of the machine learning landscape, we frame various problems in fluid mechanics as machine learning problems and we explore challenges and opportunities. We consider several relevant applications: aeroacoustic noise prediction, turbulence modelling, reduced-order modelling and forecasting, meshless integration of (partial) differential equations, super-resolution and flow control. While this list is by no means exhaustive, the presentation will provide enough concrete examples to offer perspectives on how machine learning might impact the way we do research and learn from data. 

\end{abstract}



\section{What is Machine Learning?}

Machine learning is a subset of Artificial Intelligence (AI) at the intersection of computer science, statistics, engineering, neuroscience, and biology.  The term `machine learning' was coined by \cite{Samuel1959}, to refer to \emph{the field of study that gives the computers the ability to learn without being explicitly programmed}. A more precise, engineering-oriented definition by \cite{Tom} reads: \emph{A computer program is said to learn from experience E with respect to some task T and some performance measure P, if its performance on T, as measured by P, improves with experience E}. Let us delve into this definition. 

Focusing on engineering applications, we replace the subject `computer program' with `model'. The learning of a model begins with the definition of a \emph{task} (\emph{T}): recognizing faces in images or predict flow separation from operating conditions in an airfoil, learn how to play chess or learn how to stabilize an unstable flow. These tasks can be formulated as a function from an input $\mathbf{x}\in\mathcal{X}$ to an output $\mathbf{y}\in\mathcal{Y}$. The function might define what output matches with the input (e.g. in image recognition) or what action to take when a system is at a given state (e.g. what the best next chess move is).

The second ingredient is \emph{experience} (\emph{E}), i.e. a collection of data points in $\mathcal{X}$ and $\mathcal{Y}$. These data might be available from the beginning (as in \emph{supervised learning}) or might be collected \emph{while} learning (as in \emph{reinforcement learning}). The third ingredient is an \emph{hypothesis set}, e.g. a parametric representation of the function $f\approx \tilde{f}(\mathbf{x},\mathbf{w})$ which depends on some parameters (weights) $\mathbf{w}\in\mathcal{W}$. The model $\tilde{f}(x,\mathbf{w})=w_0+w_1 x+w_2 x^2$ is an example with three parameters and rather small capacity (the set of all parabolas); an Artificial Neural Network (ANN) is an example with thousands (or millions!) parameters and a much larger capacity (potentially \emph{any} function). The set of weights, and the associated hypothesis set, define how the computer performs the task, i.e. what $\mathbf{x}-\mathbf{y}$ association it makes, or what actions $\mathbf{y}$ it takes when in state $\mathbf{x}$.

The fourth ingredient is a \emph{learning algorithm}, based on some \emph{performance measure} \emph{P}. This includes the definition of a cost function that must be minimized or a reward function that must be maximized. Learning is, in essence, an optimization problem that seeks to find the \emph{best} set of weights $\mathbf{w}$ for a given task and within a hypothesis set $f(\mathbf{x},\mathbf{w})$. The process of optimizing the weights is called \emph{training}, and can be made offline or online: in the first case, the learner is trained from a training dataset and then deployed with seldom updates afterwards; in the second case the learning is performed incrementally, from data acquired sequentially.

Finally, the last ingredient is the definition of the final hypothesis, its validation and the quantification of uncertainties. To perform unbiased evaluation of the model, not all data is used for training: a portion is retained to test the model in unseen data and, hence, to estimate its ability to generalize. Before proceeding with examples, we introduce the three paradigms of learning in the following subsections. The reader is referred to \cite{MendezBOOK,Brunton2020,Brenner2019} for reviews and perspectives on machine learning for fluid dynamics.

\subsection{Supervised Learning}
In supervised learning, a large set of data ($\mathbf{x}^*,\mathbf{y}^*$) is provided by a \emph{supervisor} and available for training. Supervised learning tasks are \emph{classification} and \emph{regression}. Classification targets the mapping of categorical data. Examples are the classification of an email as ``spam'' or ``not spam'', or the image-based classification of a flow as ``bubbly flow'' or ``churn flow''. Regression targets the mapping of continuous space. Examples are the prediction of a car's price based on mileage and age, or the prediction of the eddy viscosity given the mean flow velocity gradient in a turbulent flow.

Machine learning tools for classification and regression are entering in the literature of fluid flows. Examples of automatic data-driven classification of flow regimes are provided in \cite{Majors2018,Hobold2018,Kang2020}. Regression problem are arguably more common, encompassing surrogate model-based optimization \citep{Kim2019}, turbulence modeling \citep{Duraisamy}, non-intrusive Reduced Order Modeling \citep{Daniel2020,Hesthaven2018,Renganathan2020}, aeroacoustic noise prediction \citep{Dominique2021}, surrogate modeling \citep{Calado2023,Gkimisis2023} and system identification for prediction and control \citep{Pan2018,Brunton3932,Huang}. The main challenge in setting these problems is the choice of the hypotheses set balancing model complexity versus available data, and the implementation of physical constraints in the learning. Both aspects are briefly illustrated in Section \ref{sec:Sec2}.

Regardless of the application, the final outcome is a ``surrogate'' model that can make predictions. These models comes in various shapes and sizes: examples considered in this work are linear combination of radial basis functions (RBFS) \citep{Buhmann2003}, Artificial Neural Networks \citep{Goodfellow-et-al-2016}  or recursive expression trees as in Genetic Programming \citep{Banzhaf1997}. These models provide analytical representations, hence amenable to analytic differentiation. This enables meshless integration of Partial Differential Equations, discussed in Section \ref{sec:Sec3}.

\subsection{Unsupervised Learning}

Unsupervised learning is also known as descriptive learning. The goal is to discover patterns or characteristic features in the data. Hence the function to be learned is from the input space to itself, i.e. $f:\mathcal{X}\rightarrow \mathcal{X}$. The main unsupervised learning tasks are  \emph{dimensionality reduction} and \emph{clustering}. 

Dimensionality reduction aims at identifying a lower-dimensional representation of the data. The underlying assumption is that a few features (called \emph{hidden} or \emph{latent factors}) contain the essential information in the data. A successful face recognition algorithm, for example, focuses on image patterns that are associated with age, gender or pose, and constructs a reduced set of templates that encodes the essential information enabling recognition \citep{Swets1996}. 

Clustering aims at partitioning the data into groups (clusters) that share some common features or are similar according to certain metrics. Clustering differs from classification in its unsupervised nature: no labelled data is available, and no ``right answer'' is known upfront-- not even the number of clusters. A simple example of a clustering problem is finding customers with similar purchase behaviour as a basis for recommendation engines. Similarly, one could cluster snapshots of a flow field based on their degree of similarity and construct reduced models of a fluid flow \citep{Kaiser2014}.

Machine learning tools for clustering and dimensionality reduction are becoming increasingly popular in fluid mechanics. The quest for identifying (and objectively define) coherent structures in turbulent flows has a long history and vast literature. Linear tools such has Proper Orthogonal Decomposition (POD) have been extensively used to construct reduced order models of fluid flows \citep{Holmes}, to find optimally balanced control laws \citep{Ilak2008}, to perform correlation-based filtering \citep{Mendez_J_1} or to identify correlations between different flow quantities \citep{POD_EXTENDED_1}, to name a few examples.

Most of the decomposition methods developed in fluid mechanics are linear, and the literature has grown into a subfield of data processing often referred to as \emph{Data-Driven Modal Analysis} \citep{Taira}, where the notion of `mode' generalizes that of `coherent structure' or `principal component' or `harmonic (Fourier) mode'. Nonlinear methods of dimensionality reduction have comparatively much less history and have been popularized mostly in the last few years. Notable examples are the use of manifold learning techniques such as Locally Linear Embedding (LLE) \citep{Ehlert2019}, cluster-based reduced-order model \citep{Kaiser2014} and autoencoders \citep{Murata2019}. An illustrative application of linear and nonlinear dimensionality reduction is provided in Section \ref{sec:Sec4}.

\subsection{Reinforcement Learning}

Reinforcement learning (RL) is about learning the mapping $f:\mathcal{X}\rightarrow\mathcal{Y}$ having no data available upfront. The only viable learning approach is by trial and error. The learner is thus a decision-making \emph{agent} which interacts with an \emph{environment} by taking \emph{actions} that leads to \emph{ rewards} (or \emph{penalties}). The agent learns from mistakes and seeks to maximize the rewards (or minimize the penalties). The mapping to be learned is called \emph{policy}. This maps the state of the system to the action that needs to be taken in order to achieve the goal. The reader is referred to \cite{Arulkumaran2017,Hernandez-Leal2019} for extensive surveys, to \cite{Sutton,FrancoisLavet2018} for complete introductions to the topic and to \cite{AlexanderZai2020} for hands-on tutorials on Python.

The recent interest in this field has been motivated by standout achievements in board games \citep{Silver2016,Silver2018} and video-games \citep{Szita2012}, robotics \citep{Kober2014} and language processing \citep{Luketina2019}. An historical turning point was the success of a hybrid RL system that defeated the Korean world champion Lee Sedol in the game of Go \citep{Silver2016}. The first applications of RL in fluid mechanics were focused on the study of collective behavior of swimmers, pioneered by Koumoutsakos's group \citep{Wang2018,verma2018efficient,novati2017a,novati2019a,novati2019b}, while the first applications for flow control were presented by \cite{pivotreinforcement} and by \cite{rabault2018artificial}. \cite{bucci2019control} used RL to control the one-dimensional Kuramoto–Sivashinsky equation while \cite{beintema2020controlling} used it to control heat transport in a two-dimensional Rayleigh–Bér\-nard systems. \cite{verma2018efficient} uses RL to study how fishes can reduce energy expenditure by schooling while \cite{reddy2016learning} uses RL to analyze how bird and gliders minimize energy expenditure when soaring by exploiting turbulent fluctuations. An illustrative application of RL in flow control is provided in Section \ref{sec:Sec5}.

\section{Physics Informed Regression}
\label{sec:Sec2}

Consider the problem of finding a mapping $\mathbf{y}=f(\mathbf{x})$, with $\mathbf{x}\in\mathbb{R}^{n_x}$ and $\mathbf{y}\in\mathbb{R}^{n_y}$ using a dataset $\{\mathbf{x}_*,\mathbf{y}_*\}$ for training and a dataset $\{\mathbf{x}_{**},\mathbf{y}_{**}\}$ for validation. Let $\tilde{f}(\mathbf{x},\mathbf{w})$ denote the parametric hypothesis depending on the weights $\mathbf{w}$ and let us consider the case this is a fully connected feed forward Artificial Neural Network (ANN). These are distributed architectures consisting of a large number of simple connected units (called neurons), organized in layers \citep{Goodfellow-et-al-2016}. The mapping from input to output takes the recursive form:

\begin{equation}
\label{ANN}
    \mathbf{y}=\tilde{f}(\mathbf{x},\mathbf{w})=\sigma^{(L)}\left(\mathbf{z}^{(L-1)}\right)\quad \mbox{with}\,\begin{cases}
			\mathbf{a}^{(1)}=\mathbf{x}, \,\mathbf{a}^{(l)}=\sigma^{l}\left (\mathbf{z}^{(l)}\right) & \\
            \mathbf{z}^{(l)}=\mathbf{W}^{(l-1)}\mathbf{a}^{(l-1)}+\mathbf{b}^{(l)} & 
		 \end{cases} \, \mbox{and}\, l=1,2,\dots L\,.
\end{equation} 

Here $\mathbf{a}^{(l)},\mathbf{b}^{(l)}\in\mathbb{R}^{n_l\times 1}$ are the \emph{activation vector} and the \emph{bias vector} of layer $l$, composed of $n_l$ neurons, $\sigma^{(l)}$ is the activation function in each layer, $\mathbf{W}^{(l)}\in\mathbb{R}^{n_l\times n_{l-1}}$ is the matrix containing the weights connecting layer $l-1$ with layer $l$, and $L$ is the number of layers. The vector $\mathbf{w}\in\mathbb{R}^{n_w\times 1}$ collects all the weights and biases across the network, hence $n_w=n_L n_{L-1}+n_{L-2} n_{L-3}+\dots n_2 n_1 + n_L+n_{L-1}+\dots n_1$.

The activation functions are nonlinear functions such as hyperbolic tangents. The zoology of activation functions counts dozens of possibilities; for the scope of this talk, it suffices noticing that a sufficiently large network can `learn' approximations of \emph{any} function \citep{Cybenko1989}. Training the network simply consists in solving nonlinear least square problem which seeks to minimize a cost function. The most classic form is the mean square error $J(\mathbf{w})=||\mathbf{y}_*-f(\mathbf{x}_*,\mathbf{w})||^2_2$, with $||\bullet||_2$ denoting the $l_2$ norm.

Besides the unique function approximation capabilities of ANN, their burgeoning popularity is arguably due to two more factors. The first is the possibility to easily compute the cost function's gradient $\nabla_\mathbf{w} J(\mathbf{w})$ using the chain rule for differentiation, which leads to the popular back-propagation algorithm \citep{David}. This allows to use an arsenal of optimization tools for the training. In their stochastic `batch' formulation, these can easily handle extremely large datasets. The second is the diffusion of powerful and accessible open-source libraries such as \textit{Tensorflow}\footnote{See \url{https://www.tensorflow.org/}} or \textit{Pytorch}\footnote{See \url{https://pytorch.org/}}.

\begin{figure}[!ht]
	\centering
	\includegraphics[keepaspectratio=true,width=0.95\columnwidth]
	{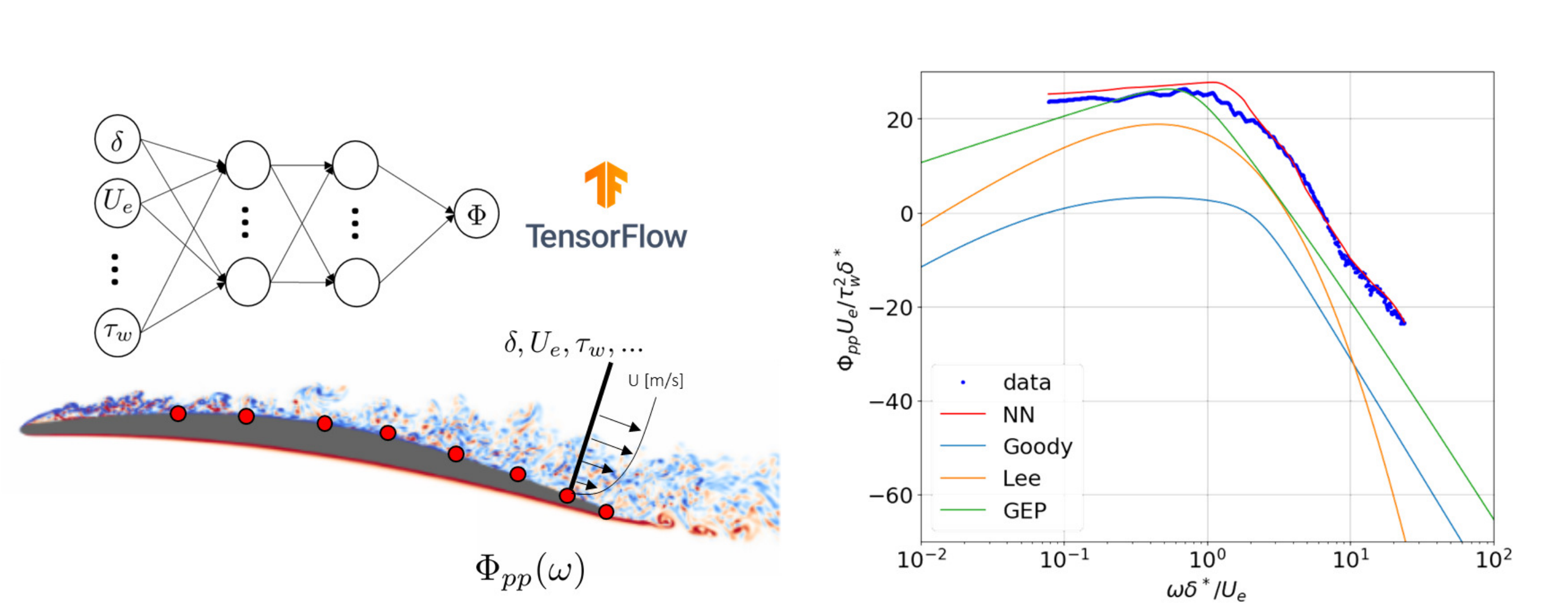}
	\caption{Wall pressure spectra prediction from from integral boundary layer parameters using an ANN trained on a large numerical and experimental database. The left figure shows the considered test case from \cite{deuse2020}. The wall pressure spectra at $x_c = 0.02$ is shown on the right and compared with the prediction of Goody's and Lee's models as well as Genetic programming and ANN models. }\label{Fig2}
\end{figure}

As first example of the regression capabilities of an ANN, we consider the problem of predicting wall pressure spectra in a turbulent boundary layer using integral parameters such as boundary layer thickness or friction coefficient. This work is presented by \cite{Joachim2}, who trained an ANN on a large dataset of numerical and experimental data, and compared the predictive performance to classic empirical correlations. These correlations are derived following several physical argument, but are ultimately adjusted using standard curve fitting tools. Figure \ref{Fig2} demonstrates the predictive capabilities of the trained ANN, consisting of three layers with ten neurons each. The input parameter consists of dimensionless numbers linking a set of carefully chosen integral parameters, so that the learned function takes the form:

\begin{equation}
    \frac{\Phi_{pp}(\omega ) U_e}{\delta^* \tau^2_w}= f\left(\frac{\omega \delta ^*}{U_e},\beta_c, R_T, C_f,H, \Delta,M , \Pi\right)\,,
\end{equation} 

\noindent where $\Phi_{pp}(\omega )$ is the wall pressure spectrum, $\delta ^*$ is the boundary layer's displacement thickness, $\tau_w$ is the wall shear stress, $U_e$ is the (local) free stream velocity, $\beta_c=(\theta/\tau_w) (dp/dx)$ is the Clauser's parameter accounting for the pressure gradient $dp/dx$ \citep{Clauser1954}, $\theta$ is the boundary layer's momentum thickness, $R_T = (\delta^*/U_e)/(\nu/u_\tau)$ is the ratio of characteristic time scales, with $\nu$ the fluid viscosity and $u_{\tau}=\sqrt{\tau_w/\rho}$ the friction velocity, $C_f = \tau_w/(0.5 \rho U_e)$ is the skin friction coefficient, $H={\delta^*}/{\theta}$ is the shape factor, $\Delta = \delta^* / \delta$, with $\delta$ the boundary layer thickness, $M=U_e/c_0$ is the Mach number with $c_0$ the free stream sound speed and $\Pi$ is Coles' wake parameter \citep{Coles1956}. The test case is the one from Deuse's DNS dataset \citep{deuse2020} and features the flow past an airfoil at Mach number $M=0.25$ and Reynolds number $Re=U_{\infty} c/\nu=1.5\times 10 ^5$. 

In Figure \ref{Fig2}, the prediction on the right is made at a location $x_c = 0.02 $ (i.e. at two \% of the chord) from the trailing edge. The pressure spectrum is compared with the two popular models of \cite{Goody2004} and \cite{Lee2018}, together with the prediction of a recursive expression trees trained via Genetic programming in \cite{Dominique2021}. Both Machine learning approaches outperform existing models in terms of predictive capabilities.

\begin{figure}[!ht]
\centering
	\includegraphics[width=0.44\textwidth]{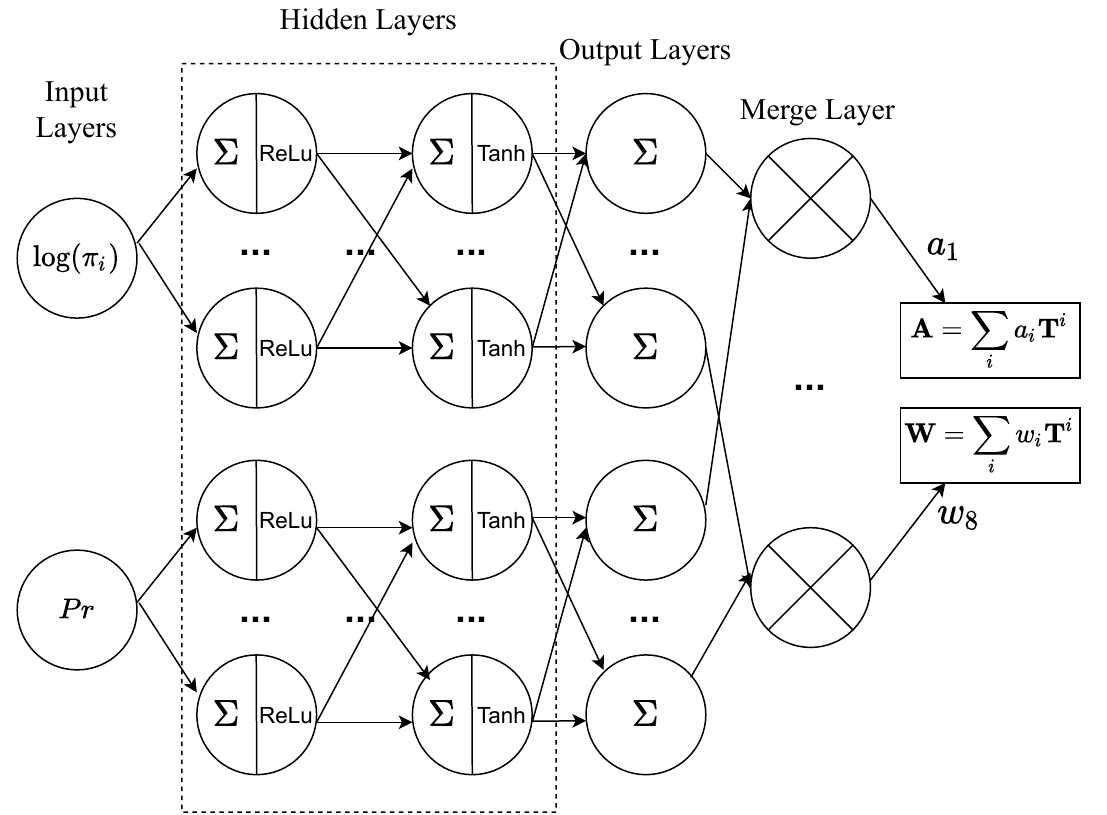}
	\includegraphics[width=0.45\textwidth]{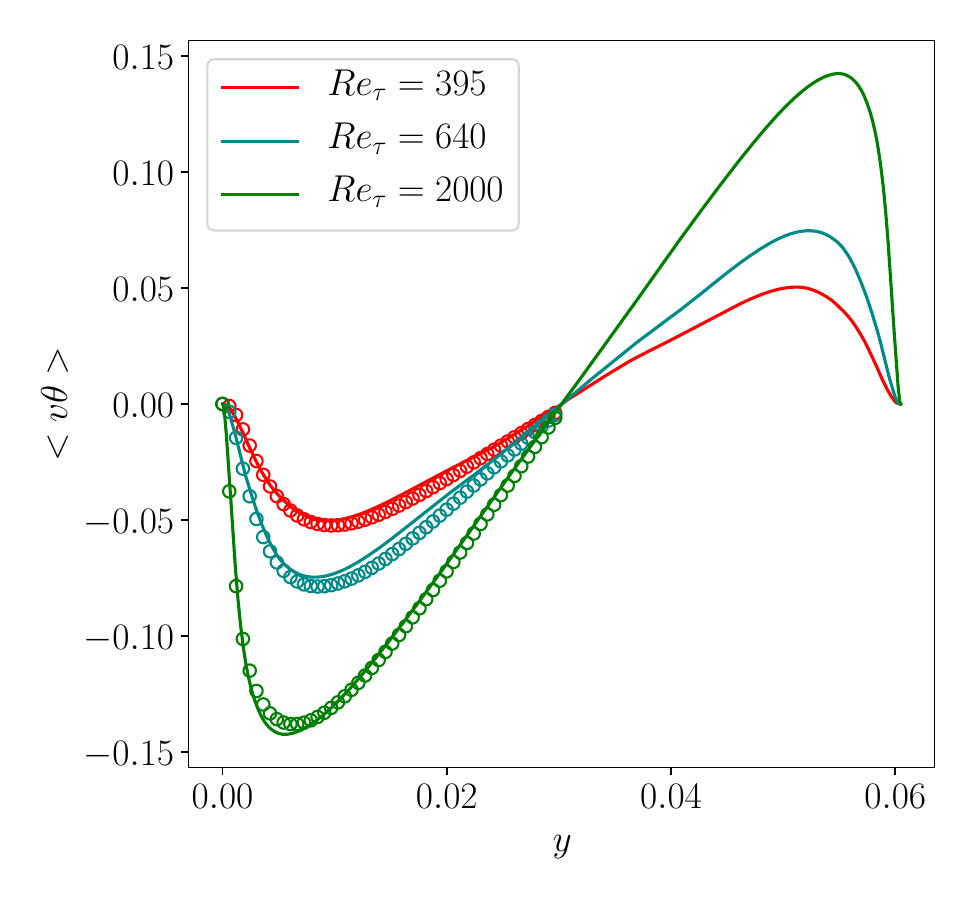}
	\caption{Thermal Turbulence Modeling via ANN. The left figure shows the scheme of the ANN used to predict the coefficients $a_i$ and $w_i$ in \eqref{Matilde_EQ}. The right figure depicts the wall-normal turbulent heat flux resulting from the OpenFoam simulations carried out with the data-driven thermal model in case of turbulence channel flow at $Pr=0.025$. Images adapted from \cite{Fiore2}.}\label{Fig3}
\end{figure}

Physical consideration can be added during the training process using two possible approaches: either by defining an hypothesis that respects certain physical constraints, either by using constraints or penalization in the optimization driving the training. Both methods are illustrated by \cite{Fiore1,Fiore2} for the prediction of turbulent heat flux in low Prandtl number fluids. In this context, the mapping to be learned is an algebraic model for the turbulent heat flux $\overline{\mathbf{u}\theta}$, based on a generalized gradient hypothesis. The hypothesis set is designed to be of the form:

\begin{equation}
    \label{Matilde_EQ}
     \overline{\mathbf{u} \theta}=-\mathbf{D} \nabla T \quad \mbox{with}\quad \mathbf{D}=\left[(\mathbf{A}+\mathbf{A}^T) (\mathbf{A}^T+\mathbf{A}) + \frac{k}{\sqrt{\varepsilon}} (\mathbf{W}-\mathbf{W}^T)\right]\,,
\end{equation} where $T$ is the local average temperature, $k$ and $\varepsilon$ are the turbulent kinetic energy and dissipation rates, and the tensors $\mathbf{A}$ and $\mathbf{W}$ are written as linear combinations of basis tensors $\mathbf{T}_i$:

\begin{equation}
    \label{Matilde_EQ_2}
     \mathbf{A}=\sum_{i=1}^n a_i \mathbf{T}_i \,\,\mbox{with}\,\, a_i = f_i (\pi_i, Re_\tau, Pr)\quad \mbox{and}\quad \mathbf{W}=\sum_{i=1}^n w_i \mathbf{T}_i \,\,\mbox{with}\,\, w_i = g_i (\pi_i, Re_\tau, Pr)\,.
\end{equation}

The coefficients $a_i$ and $g_i$ are functions of the turbulent Reynolds number $Re_\tau$, the Prandtl number $Pr$ and a set of invariant basis $\pi_i$, i.e. isotropic quantities that are invariant under rotation of the coordinate system. The functions $f_i$ and $g_i$ for the coefficients in \eqref{Matilde_EQ_2} are needs to be learned by an ANN.

Neither the invariant bases $\pi_i$ nor the basis tensors $\mathbf{T}_i$ are shown in this short article and the reader is referred to \cite{Fiore1,Fiore2} for more details. For the purposes of this overview, it suffices noticing that \eqref{Matilde_EQ} and \eqref{Matilde_EQ_2} limits the margin of manoeuvre of the ANN in making predictions that satisfy the Galilean and geometrical invariant properties as well as rotational invariant. Moreover, because $\mathbf{D}$ is at least positive semi-definite, it can be shown that the predicted heat flux respects the second principle of thermodynamics and only flows from higher temperatures to lower ones.
Figure \ref{Fig3} demonstrates the predictive capabilities of the proposed ANN for  turbulent channel flows at different Reynolds numbers and $Pr=0.025$. The network was trained up to  $Re_{\tau}=640$. Hence, the model seems able to accurately extrapolate the turbulent heat flux at higher Reynolds numbers ($Re_{\tau}=2000$).

\section{From Regression to Super-resolution and Meshless PDE Integration}
\label{sec:Sec3}

Most of the parametrized models $\tilde{f}(\mathbf{x},\mathbf{w})$ employed in machine learning are easily differentiable with respect to the inputs. It is thus natural to further constrain the training by imposing that $\tilde{f}$ is solution of a Partial Differential Equation (PDE). Given a differential operator $\mathcal{N}\{u\}$ (for example $\mathcal{N}\{u(x,t)\}=\partial_t u+ \partial_x u$ for the 1D advection equation) and $\mathcal{N}\{u\}=0$ a PDE, inserting the ansatz $u\approx \tilde{f}(\mathbf{x},\mathbf{w})$ returns a least square problem in $\mathbf{w}$, whose solution gives an \emph{analytic} approximation of the PDE solution.

No computational mesh is needed. On the contrary, the resulting approximation can be evaluated at \emph{any} mesh, hence enabling super-resolution (a term often borrowed from computer vision \citep{Bashir2021}) when the training data comes from experiments. 

The idea of meshless integration of PDEs using ANN is rather old (see \cite{Lagaris1998}), and the same is true for the PDE integration via RBFs (see \cite{Fornberg2015}). Nevertheless, recent contributions on Physics Informed Neural Networks (PINs \cite{Raissi2019}) and brilliant online tutorials available online\footnote{See \url{https://maziarraissi.github.io/research/1_physics_informed_neural_networks/}.} are making these tools more accessible.

This talk shows an example of data-driven meshless integration of the Poisson equation to measure pressure fields from image velocimetry. This work is presented by \cite{Sperotto} and uses Gaussian Radial Basis Functions as function approximators. These are linear models (w.r.t. to the weights $\mathbf{w}$) of the form

\begin{equation}
\label{RBF_G}
    \mathbf{y}=\tilde{f}(\mathbf{x},\mathbf{w})=\sum^{n_{\phi}}_{k=1}w_k \varphi_k(\mathbf{x}| \mathbf{x}^*,c_k)\quad \mbox{with} \quad \varphi_{k}(\mathbf{x}| \mathbf{x}_k^*, c_k)=\exp \left(-c_k^{2}\left\|\boldsymbol{x}-\boldsymbol{x}_{\boldsymbol{k}}^{*}\right\|^{2}\right).
\end{equation} 

Here $c_k>0$ are the {\em shape parameters} and $\boldsymbol{x^*_k}$ are the {\em collocation points} of the basis functions. Derivatives of these expansions are trivially computed by replacing the basis functions with their derivatives. 

In the pressure integration algorithm, the RBFs are first used to construct an analytic expression of the velocity field $\mathbf{u}(\mathbf{x})$ from data produced by Particle Image Velocimetry (PIV) or Lagrangian Particle Tracking (LPT). This step is cast in the form of a constrained least square problem ensuring that the approximation respects boundary conditions and physical priors such as the divergence free condition for an incompressible flow. The resulting velocity approximation is then introduced in the pressure Poisson equation, leading to a second least square problem. For an incompressible flow, this problem reads:

\begin{equation}
\label{RBF_P}
    \mbox{Given} \,\, p=\sum^{n_{\phi}}_{k=1}w_k \varphi_k(\mathbf{x}| \mathbf{x}^*,c_k) \, \rightarrow \, \nabla^2 p=\sum^{n_{\phi}}_{k=1}w_k \nabla^2\varphi_k(\mathbf{x}| \mathbf{x}^*,c_k)=-\rho \nabla \cdot (\mathbf{u} \nabla \mathbf{u})
\end{equation}

This linear problem is solved using the same techniques used for the velocity approximation, with constraints imposing boundary conditions for the pressure field.

\begin{figure}[!ht]
\centering
	\includegraphics[keepaspectratio=true,width=0.47\columnwidth]
	{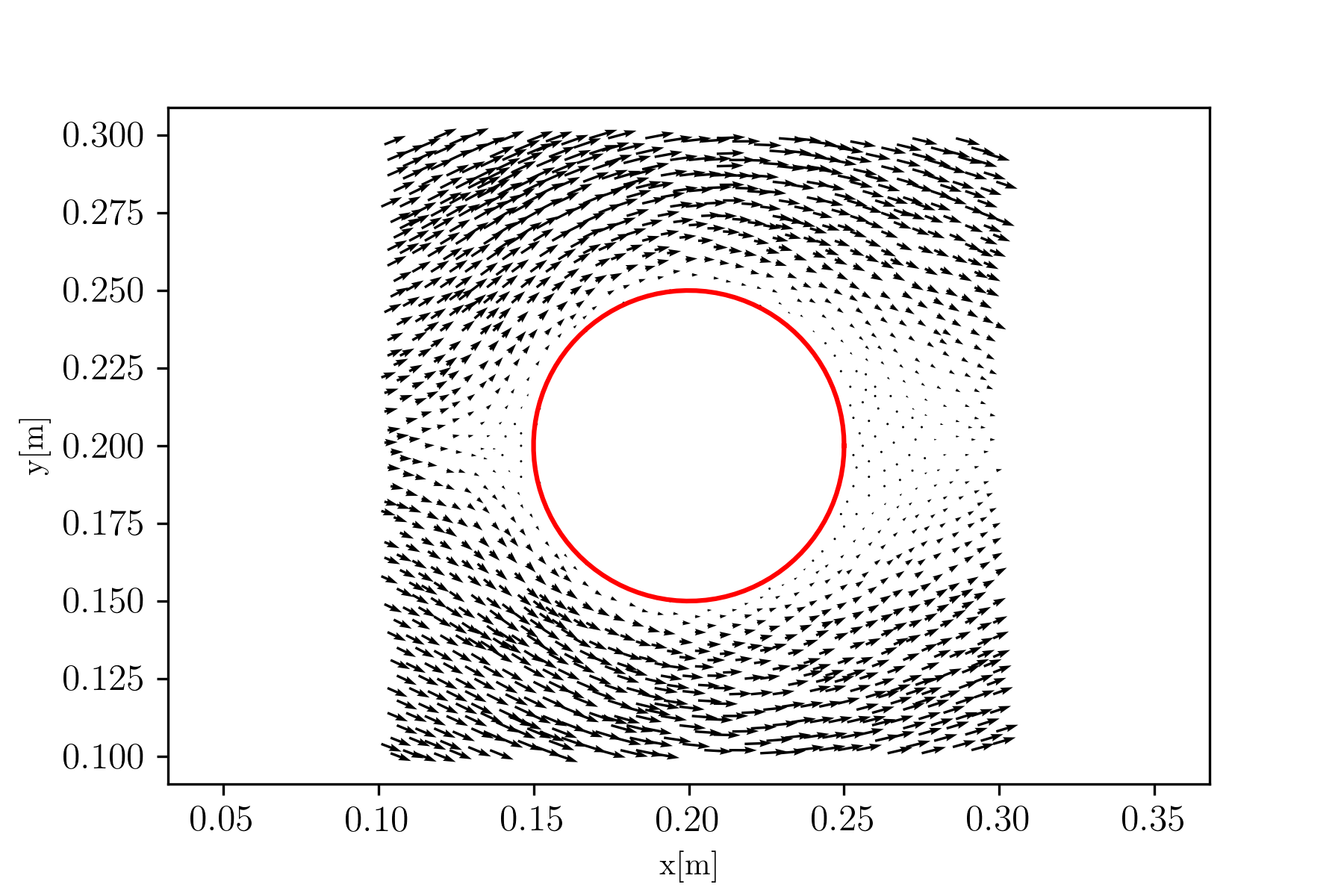}
	\includegraphics[keepaspectratio=true,width=0.47\columnwidth]
	{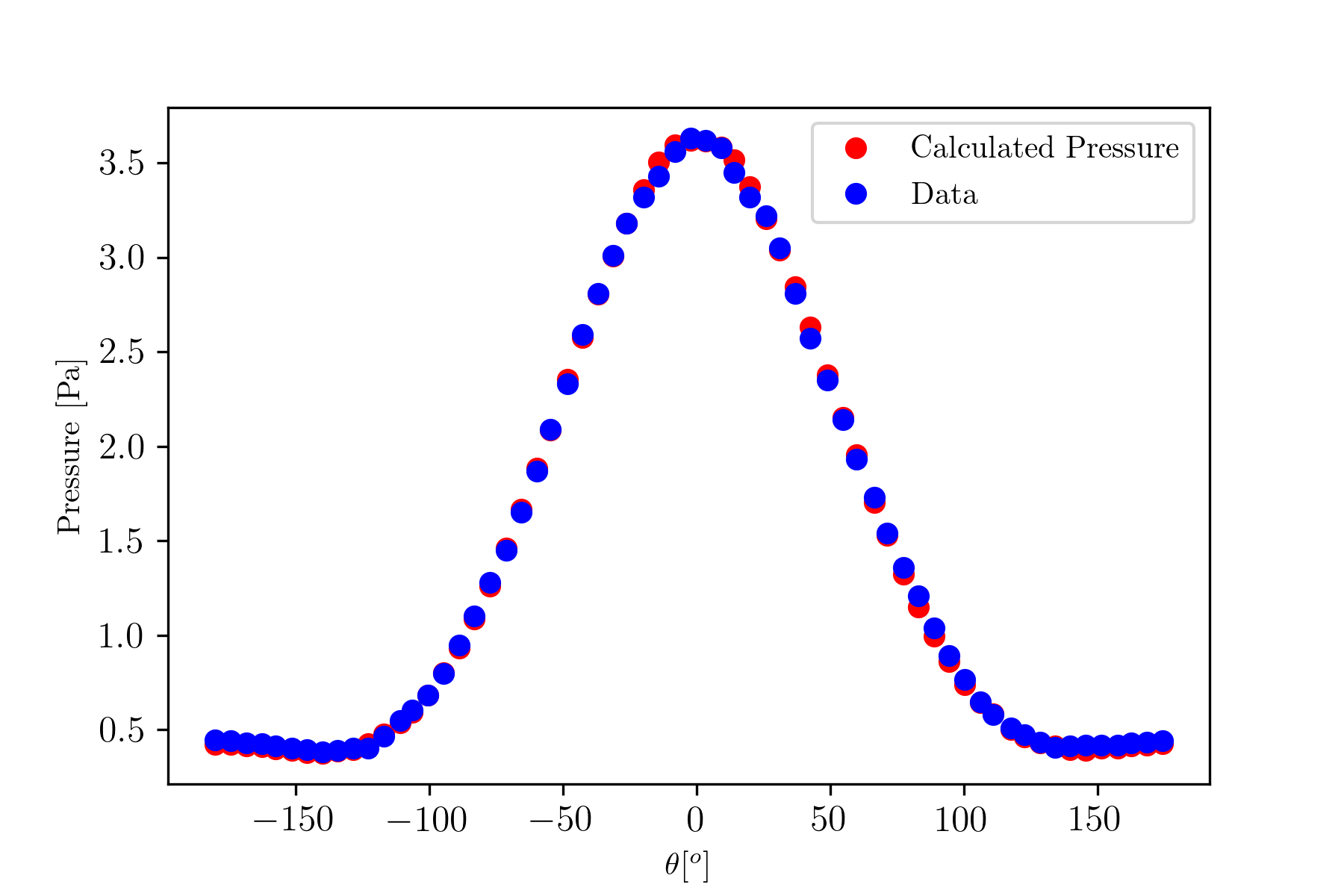}
	\caption{Data-driven pressure reconstruction from 2D scattered velocity fields. The left figure shows a snapshot of the scattered velocity data in the flow past a cylinder (dataset re-sampled from CFD simulations in \cite{Rao2020}). The right figure compares the  meshless RBF integration (red circles) to the CFD data (blue circles), along the cylinder surface.}\label{Fig4}
\end{figure}

An illustrative result is shown in Figure \ref{Fig4}. This is the pressure integration in the flow past a cylinder in laminar conditions. The figure on the left is a snapshot of the scattered velocity field from which the pressure field is computed via RBFs. The velocity data is taken by down-sampling a CFD simulation by \cite{Rao2020} so as to obtain a scattered field consisting of $n_s=18755$ vectors, simulating a LPT measurements. The figure on the right compares the resulting pressure distribution along the cylinder, as a function of the angular coordinate $\theta$, with the CFD results. The reconstruction is in excellent agreement.

\section{From Modal Analysis to Manifold Learning}
\label{sec:Sec4}

The quest for identifying coherent structures in turbulent flows has a long history in fluid mechanics (see \cite{Holmes,Hussain1983}). The most ubiquitous tool developed at the scope is the Proper Orthogonal Decomposition (POD), which is essentially equivalent to the Principal Component Analysis (PCA) in the machine learning literature \cite{Bishop2006}. This is a linear tool: it decomposes a dataset as a linear combination of contributions called modes. These are computed to minimize the error in an approximation built excluding some of the modes.

Many variant exist (see \cite{Mendez2019,Mendez2023}) and the range of application spans filtering, compression, feature identification, reduced order modeling and more. However, linear tools are only a small subset in the arsenal of data compression and manifold learning tools available to the machine learner \citep{Bishop2006}. 
We illustrate the use of a nonlinear reduced order modeling tool known as kernel (kPCA) \citep{Schoelkopf1997}. This is essentially a \emph{kernelized} version of the POD. The underlying idea is to perform the PCA/POD on a dataset which has been first transformed by a nonlinear function called kernel function $\xi$. We briefly illustrate the underpinnings of the kPCA starting form PCA.

In a linear decomposition, each mode represents the component of the dataset along a certain \emph{basis element}. Assume that the dataset is reshaped into a matrix $\mathbf{X}\in \mathbb{R}^{n_s\times n_t}$, collecting the snapshots of $n_s$ spatial points along its columns and the temporal evolution (time series of length $n_t$) in each point along its rows. The PCA seeks the optimal basis from the eigenvalues of $\mathbf{X} \mathbf{X}^T$.
An approximation $\tilde{\mathbf{X}}$ is computed as:

\begin{equation}
	\label{PCA}
	\tilde{\mathbf{X}}(x_i,t_k)=\sum^{R}_{r=1} c_r (t_k) \phi_r(x_i) \,\, \mbox{where}\,\, c_r (t_k)=\phi^T_r(x_i) \mathbf{X}(x_i,t_k)=\mathbf{z}^T_r[k],
\end{equation} with $(\mathbf{X} \mathbf{X}^T)\phi_r=\lambda_r \phi_r$. Here $x_i$ and $t_k$ are, respectively, the spatial and temporal meshes. Both the basis elements (principal components) $\phi_r(x_i)$ and the coefficients $c_r (t_k)=\mathbf{z}^T_r$ are stored as column vectors $\phi_r\in \mathbb{R}^{n_s\times 1}$ and $\mathbf{z}_r\in\mathbb{R}^{n_t\times 1}$. These coefficients are the projection of the data onto the basis vectors. Each mode is a field $\phi_r(x_i)$ which evolves in time as $c_r (t_k)$. By truncating the expansion to $R\ll \mbox{rank}(\mathbf{X})$, a low order representation is built.

In kPCA, we construct the basis from the eigenvectors of the kernelized matrix $\xi (\mathbf{X})\in\mathbb{R}^{n_F\times n_t}$. This represents a nonlinear mapping onto a space, called the \emph{feature space}, of dimension $n_F$ (possibly infinite). Linear operations in the feature space (e.g. projections) are nonlinear in the original space. The principal components $\phi_{\xi r}\in\mathbb{R}^{n_F\times 1}$ (i.e. the eigenvectors of $\xi(\mathbf{X})\xi^T(\mathbf{X})$) live in the feature space and are nonlinear functions which offer more model capacity than the linear expansion in \eqref{PCA}. 

In practice, we might not even need an explicit definition of $\xi$ because we are solely concerned with inner products in the feature space. We can take these using the \emph{kernel trick}\citep{Bishop2006,Schoelkopf1997}. This allows for avoiding operations in the feature space and for computing the projection of the data $\mathbf{X}$ onto the principal components of the feature space $\phi_{\xi r}$ without ever computing $\phi_{\xi r}$. The trick goes as follows.

Let us write the $\phi_{\xi r}$ as linear combinations of the features:

\begin{equation}
\label{eq8}
\phi_{\xi r}=\sum^{n_p}_{i=1}a_{r}(t_k)\xi(\mathbf{X}(x_i,t_k))=\xi(\mathbf{X}) \mathbf{a}_r\,.
\end{equation}

We introduce this ansatz into the eigenvalue problem, which we multiply by $\xi (\mathbf{X})^T$:

\begin{equation}
\label{eq9}
\xi(\mathbf{X})^T(\xi(\mathbf{X})\xi(\mathbf{X})^T) \xi(\mathbf{X}) \mathbf{a}_r =\lambda_r \xi(\mathbf{X})^T \xi(\mathbf{X}) \mathbf{a}_r\rightarrow \mathbf{K}^2\mathbf{a}_r
=\lambda_r \mathbf{K}\mathbf{a}_r\,.
\end{equation} where $\mathbf{K}=\xi(\mathbf{X})^T\xi(\mathbf{X})$ collects the inner products in the feature space. If the kernel function is carefully chosen \citep{Bishop2006}, it is possible to compute these inner products via a \emph{kernel function}:

\begin{equation}
\label{eq10}
\mathbf{K}_{i,j}=\xi(\mathbf{x}_i)^T\xi(\mathbf{x}_j)=\kappa(\mathbf{x}_i,\mathbf{x}_j)\,,
\end{equation} where $\mathbf{x}_i$ and $\mathbf{x_j}$ are two snapshots of the data.
If this matrix is invertible, equation \eqref{eq9} reduces to $\mathbf{K}\mathbf{a}_r
=\lambda_r\mathbf{a}_r$, i.e. an eigenvalue problem for $\mathbf{K}$. Given the eigenvectors $\mathbf{a}_r$, the projection of a feature vector $\xi(\mathbf{x}_i)$ onto $\phi_{\xi r}$ in the feature space gives:

\begin{equation}
\label{eq11}
\mathbf{z}^T_r=\phi^T_{\xi r} \xi(\mathbf{X})=\mathbf{a}^T_r\xi(\mathbf{X})^T\xi(\mathbf{X})=\mathbf{a}^T_r\kappa(\mathbf{X},\mathbf{X})\,. 
\end{equation} 

\begin{figure}[!ht]
	\centering
	PCA (Global Root Mean Square Error: 0.00022869072551777843)\\	 
		\includegraphics[keepaspectratio=true,width=0.35 \columnwidth]
	{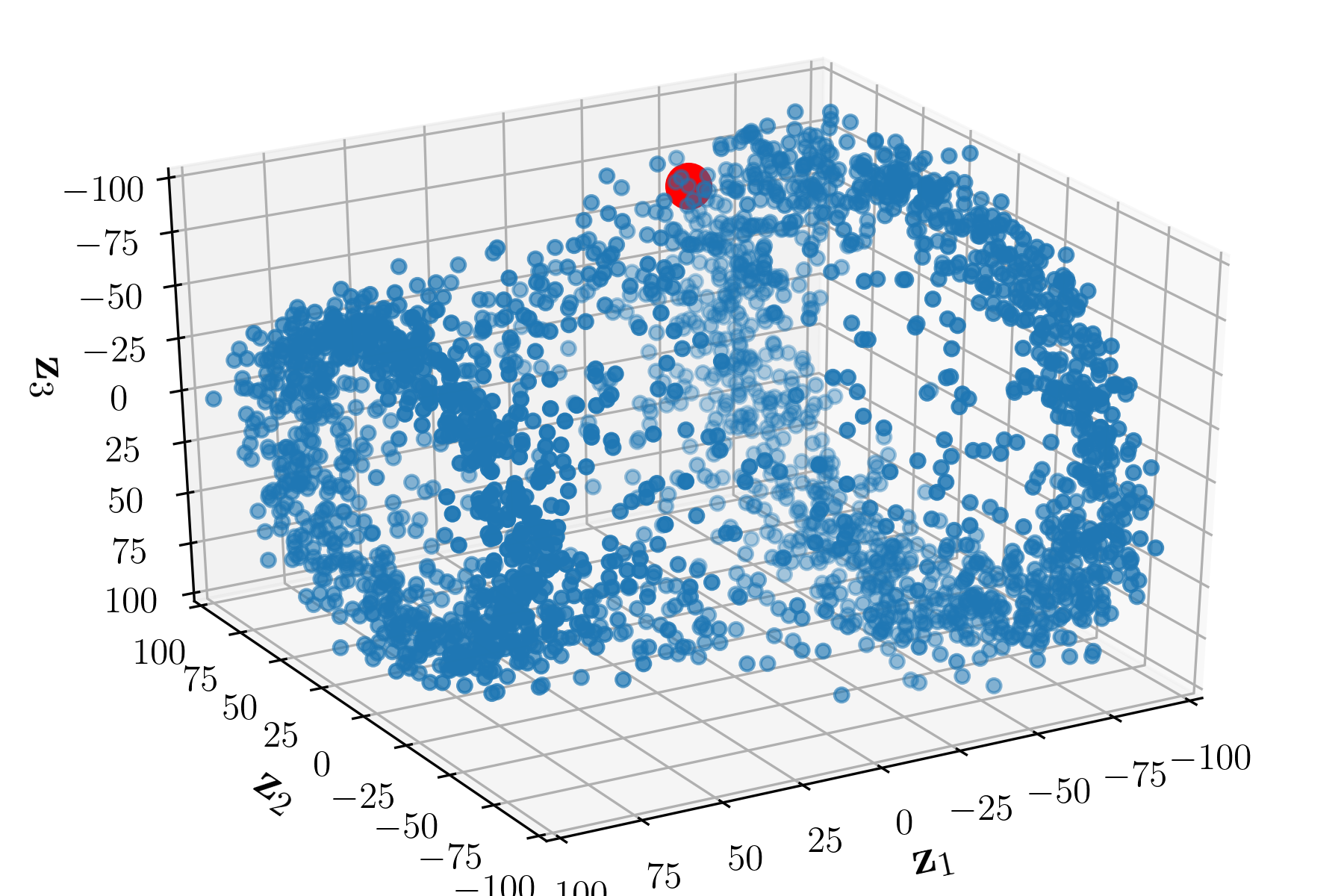}
	\includegraphics[keepaspectratio=true,width=0.5 \columnwidth]
	{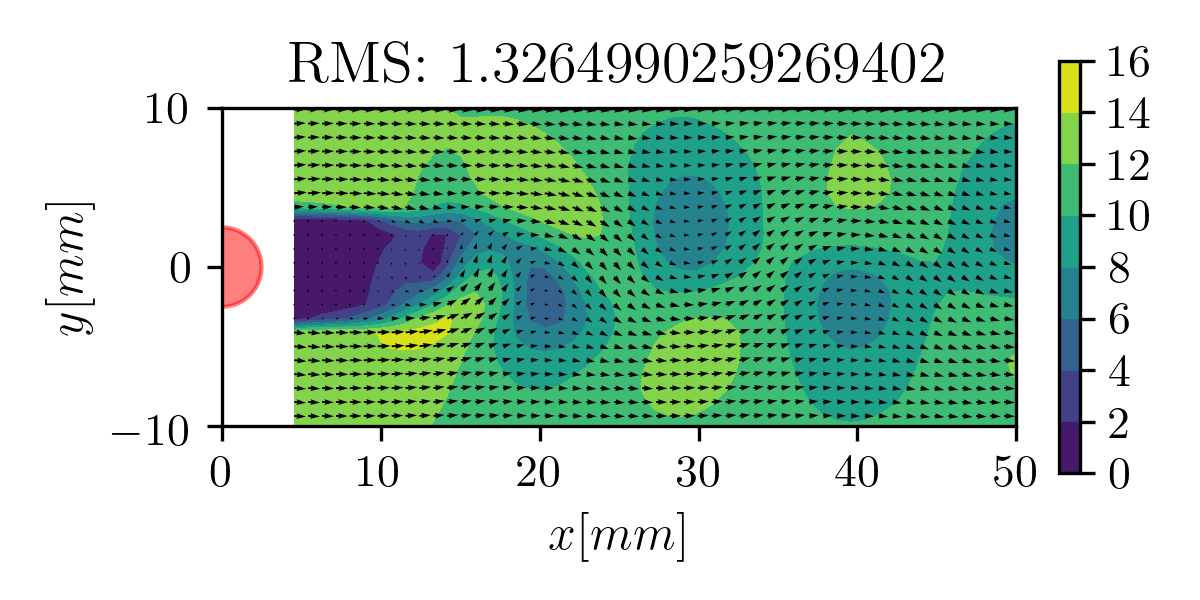}\\
	\vspace{2mm}	
		KPCA (Global Root Mean Square Error:  3.385300114380556e-17) \\
	\includegraphics[keepaspectratio=true,width=0.35 \columnwidth]
	{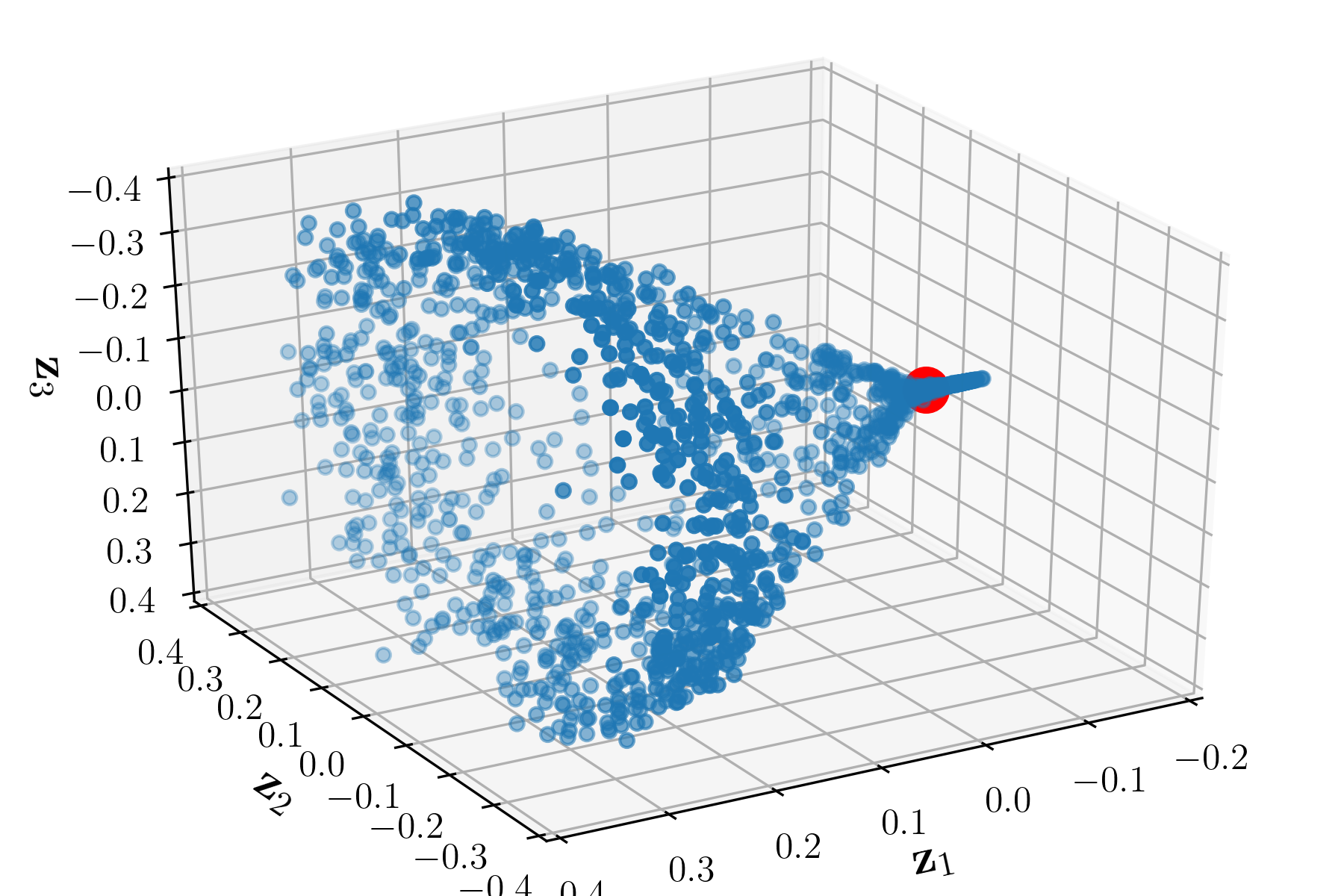}
	\includegraphics[keepaspectratio=true,width=0.5\columnwidth]
	{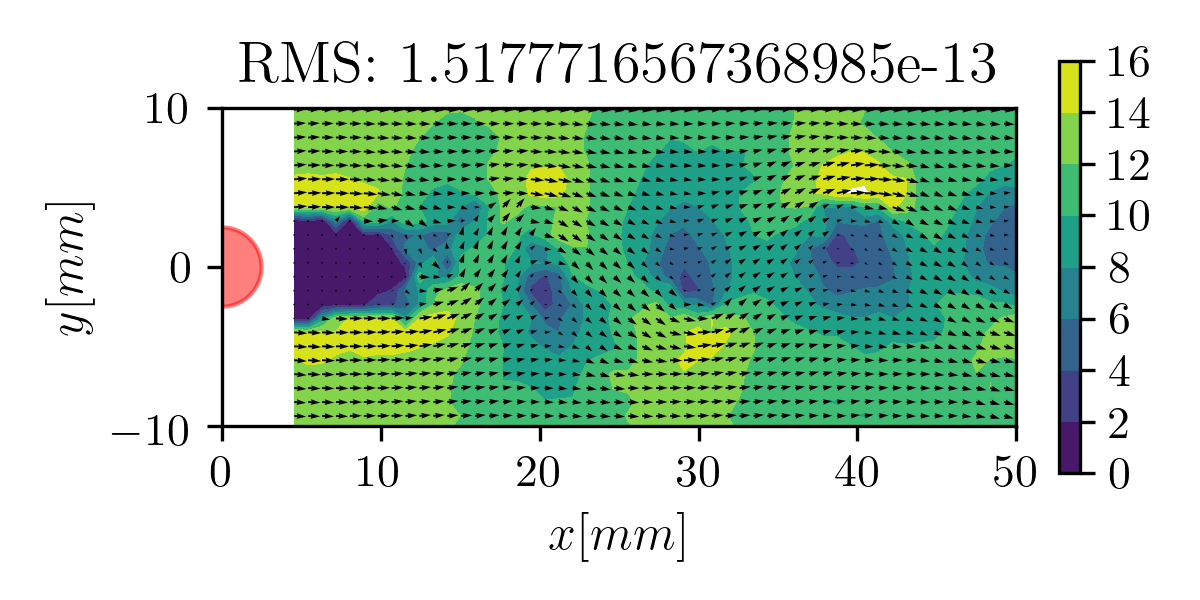}\\
	\caption
	{Linear and nonlinear dimensionality reduction via PCA and kPCA. The test case is the flow past a cylinder analyzed via TR-PIV in \cite{Mendez2020}. The 3D scatter plots on the left shows the manifold identified projecting of the fields or the kernelized fields onto the first three PCA and kPCA components. The red points on these scatter plot correspond to the snapshot shown on the right. The global and instantaneous RMS are indicated for both approximation.}\label{figFin}
\end{figure}

Reconstructing an approximation of the data from a few set of kPCA components is non trivial problem and its presentation is here omitted (see \cite{Mendez2023a}). Interested readers are referred to \cite{ikPA}. We close this section with an application, illustrated in Figure \ref{figFin}.

The dataset consists of $n_t=13200$ velocity fields of the flow past a cylinder of diameter $d=5\mbox{mm}$, measured via Time Resolved Particle Image Velocimetry (TR-PIV) at $f_s=3kHz$ over a grid of $71\times30$ points. Details on the experimental work are provided in \cite{Mendez2020}. This test case is characterized by a large scale variation of the free-stream velocity, which is decreased from a first steady state at $U_\infty\approx12\mbox{m/s}$ to second steady state at $U_\infty\approx8\mbox{m/s}$. The figures on the right show an approximation of a velocity field snapshot using the three leading modes from the PCA (top) and from the kPCA (bottom). The velocity fields are snapshot vectors $\mathbf{x}_i$ in a $n_s=2\times71\times30=4260$ dimensional space. 

The trajectory of the dataset along the three leading modes is shown on the left for both the PCA (top) and the kPCA (bottom). The manifolds are quite different: in the PCA, the learned 3D representation collapse around a paraboloid with principal axis along $\mathbf{z}_3$. Each of the steady states corresponds to circles at constant $\mathbf{z}_1$, and the transient leads the shift from the first circular orbit (with larger radius) to the second one. The existence of such paraboloid in an optimally linear basis of eigenfunctions was firstly derived in \cite{NOACK2003} for much lower Reynolds numbers. The 3D mapping produced by the kernel PCA appears as a distorted version of the one produced by the PCA. The kinematics of the data in this space is similar to the PCA, but the region of higher velocities are nonlinearly compressed in orbits of much lower radius. 

The global (i.e. over both space and time) root mean square (RMS) error is reported on the top of each figure while the local (i.e. on the shown snapshot) RMS is shown on the title of each contour plot. While the PCA does an excellent job in approximating the flow field with only three modes, the kPCA approximation is perfect, with the global RMS hitting machine precision. In other words, all the information contained in the trajectory of a system in $\mathbb{R}^{4260}$ has been mapped onto $\mathbb{R}^{3}$. Time series forecasting and modeling is much easier in $\mathbb{R}^{3}$ than in $\mathbb{R}^{4260}$.

\section{From Optimal Flow Control to Reinforcement Learning}
\label{sec:Sec5}

Flow control aims at interacting and manipulating flow systems to improve their engineering benefits. This is essential in countless applications, including drag or noise reduction, optimal wind energy extraction, stability of combustion systems or flight, and more \citep{el-Hak2000}. In its most abstract formulation, the (flow) control problem is essentially a functional optimization problem constrained by the (fluid) systems’ dynamics\citep{stengel1994optimal}. The goal is to optimize a function that measures the controller performances, i.e. its ability to keep the flow (the plant or the environment) close to the desired state (or dynamics). Although solid theoretical tools are available from optimal control theory \citep{stengel1994optimal}, the analytic derivation of these optimal laws is a formidable endeavour. To be practical, it is currently out of the reach for the most relevant problems in fluid mechanics.

Recent advances in machine learning are giving prominence to the so-called ``black-box" or ``model-free" paradigm, whereby an algorithm learns the optimal control action by trial and error. A comprehensive survey of the most popular machine learning tools for this scope is presented by \cite{Pino}. This article presents a synopsis of \cite{Pino}, considering only one algorithm from Reinforcement Learning and one simple illustrative test case: the problem of cancelling waves in a 1D advection equation.

\begin{figure}[!ht]
\centering
	\includegraphics[keepaspectratio=true,width=0.88\columnwidth]
	{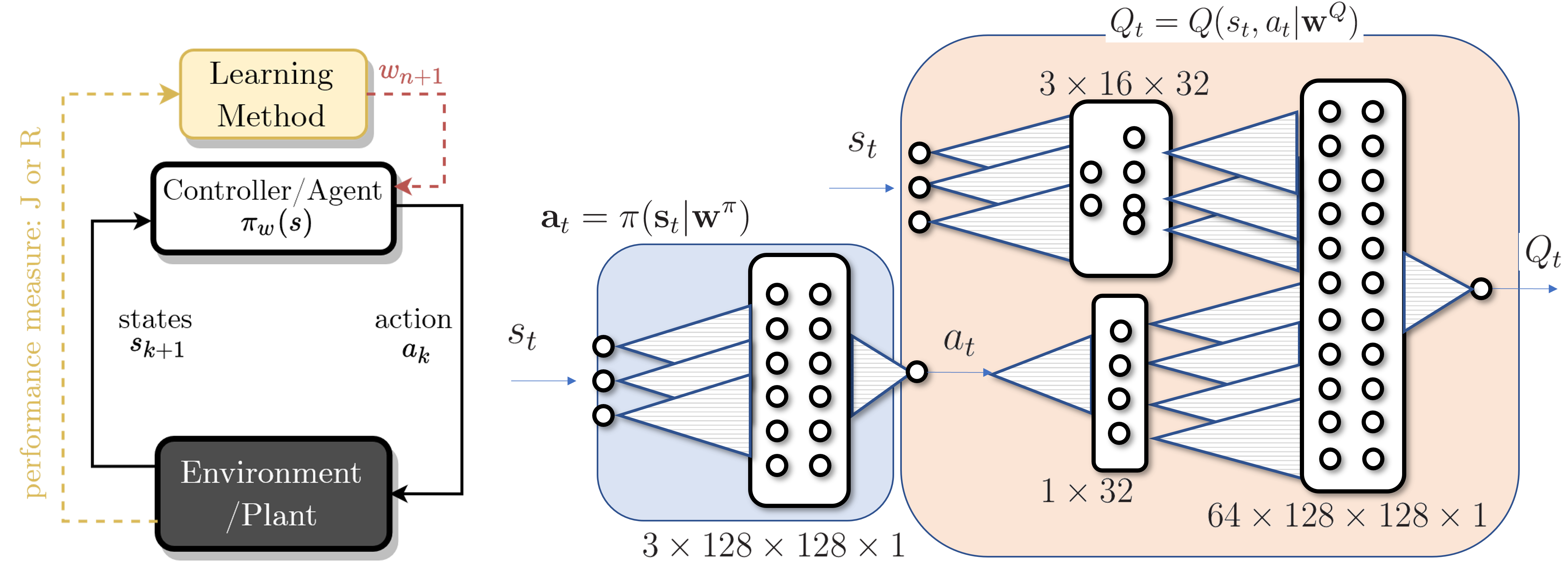}
	\caption{Application of Reinforcement Learning for flow control. The figure on the left shows the general idea: a learning method calibrate a controller which interacts with an environment via actions and collects rewards measuring its closeness to the objective. On the right: ANN architectures used in the DDPG implementation illustrated in this work. The network $\pi$ (actor) parametrizes the state-action policy while the network $\pi$ (critic) learns to estimate the value of each state-action pair.}\label{Fig5}
\end{figure}

The general framework of machine learning control methods is shown in Figure \ref{Fig5}, on the left. One aims at controlling a dynamical system in an episodic framework. Within an episode ($\mbox{ep}$) of duration $T$, we interact with the system $N=T/\Delta t$ times, at regular intervals $\Delta t$, by performing a sequence of actions $\mathbf{a}_1,\mathbf{a}_2\dots \mathbf{a}_N$ and observing the sequence of states $\mathbf{s}_1,\mathbf{s}_2\dots \mathbf{s}_N$. We denote as $\mathbf{s}_k\in\mathbb{R}^{n_s}$ and $\mathbf{a}_k\in\mathbb{R}^{n_a}$ the state vector of the system and the action at the time step $t_k=k\Delta t$. 
The agent (aka controller) must learn a policy $\mathbf{a}_t=\mathbf{\pi}(\mathbf{s}_k)$. This can be deterministic or stochastic, in which case the parametrization returns the parameters of a distribution from which the action is sampled.




We define a reward function $r_k=r(\mathbf{a_k},\mathbf{s}_k)\in\mathbb{R}$ which measures the control performances and which is large when the agent is close to the objectives. At the end of each episode, the agent collects a cumulative reward $R=\sum^{N}_{k=1}\gamma^{k}r(\mathbf{a_k},\mathbf{s}_k)$, with $\gamma \in[0,1]$ a discount factor which prioritize immediate rewards. We give a value $V^{\pi}(\mathbf{s}_k)$ to a state as the cumulative rewards achievable from $k$ to $N$ following a certain policy. This is called \emph{value function}. Similarly, we give a value for the state-action pair $Q^{\pi}(\tilde{\mathbf{a}}_k,\mathbf{s}_k) $, as the reward achievable if an action $\tilde{\mathbf{a}}_k$ is taken at the iteration $k$ and the policy is followed for the rest of the episode. This is called \emph{Q-function}. Both functions can be written in a recursive form. Assuming that the environment is stochastic (usually modeled as Markov Decision Process), the recursive form of the Q-function reads:

\begin{equation}
    \label{Belleman}
    Q^\pi(\mathbf{s}_t,\tilde{\mathbf{a}}_t)=
    \mathbb{E}_{\sim \mbox{ep}} \biggl [ r(\tilde{\mathbf{a_k}},\mathbf{s}_k)+\sum^N_{k=t+1}\gamma^{k-t} r({\mathbf{a_k}},\mathbf{s}_k)\biggr]=\mathbb{E}_{\sim \mbox{ep}} \biggl [r(\tilde{\mathbf{a_k}},\mathbf{s}_k)+ \gamma Q^\pi(\mathbf{s}_t,\tilde{\mathbf{a}}_t)\biggr]\,
\end{equation} where $\mathbb{E}_{\sim \mbox{ep}}$ is the expected value computed along the various episodes. This recursion is known as Bellman equation and is the foundation of dynamic programming.

We now have two paths to derive the optimal policy. In the so called \emph{actor-only}, we seek to find the policy that maximized the cumulative rewards. That is, we look for the weights, in the parametric function $\mathbf{a}_t=\mathbf{\pi}(\mathbf{s}_k,\mathbf{w}^{\pi})$, that maximize the value of the initial state $V^{\pi}(\mathbf{s}_1)= \mathbb{E}_{\sim \mbox{ep}} [r(\tilde{{a_k}},\mathbf{s}_k)+\gamma V^{\pi}(\mathbf{s}_2)]$.  In the so called \emph{critic-only}, we take an indirect path: we focus on learning the value of each state and each action-state pair. We might thus use \eqref{Belleman} to train a parametric function of the form $Q_t=Q(\mathbf{s}_t,\mathbf{a}_t,\mathbf{w}^{Q})$. If this correctly estimates the Q-function, then the best action to take is simply the greedy $\mathbf{a}_t=\mbox{argmax}_a Q(\mathbf{s}_t,\mathbf{a})$.

Modern RL algorithms are  \emph{actor-critic}, i.e. combine both approaches and employ two ANNs: one for learning the policy and the other to learn the Q-function. In this work, we illustrate the application of Deep Deterministic Policy Gradient (DDPG), by \cite{Lillicrap2015}, which aims at learning a deterministic policy in an actor-critic framework. The essence of the algorithm is the calculation of the gradient $\nabla _\mathbf{w} V^{\pi}(\mathbf{s}_1)$ which is computed in batches of $M$ interactions and reads:

\begin{equation}
    \label{Update_DDPG}
    \nabla _\mathbf{w} V^{\pi}(\mathbf{s}_1)\approx \frac{1}{M} \sum_{i} \nabla_a Q\bigl(\mathbf{s}_t,\mathbf{a}_t=\pi(\mathbf{s}_t,\mathbf{w}^{\pi}),\mathbf{w}^{Q}\bigr)\,\nabla_\mathbf{w}^{\pi}\pi(\mathbf{s}_t,\mathbf{w}^{\pi})\,,
\end{equation} where $i$ is the index scanning the batch of interactions taken. The algorithm contains several other tricks, such as the implementation of a replay buffer collecting previous experience and the use of a sort of under-relaxation when updating the weights of both the policy and the Q networks. Details can be found in \cite{Lillicrap2015}.

\begin{figure}[!ht]
\centering
	\includegraphics[keepaspectratio=true,width=0.8\columnwidth]
	{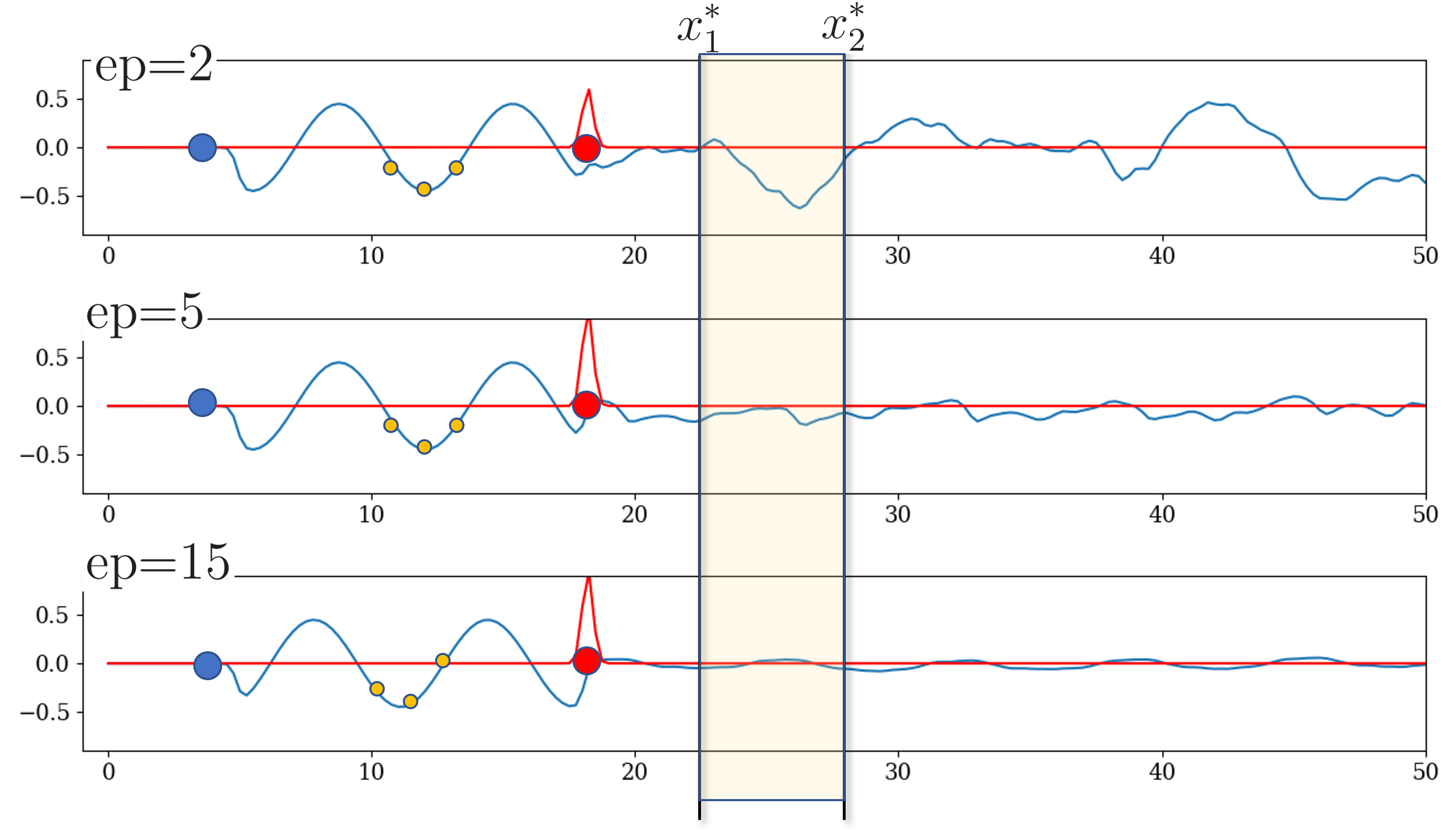}
	\caption{Results of the DDG implementation for a wave cancelling problem. Waves are produced at $x=5$ (blue marker) and travel downstream $x\rightarrow \infty$. A controller is located at $x=18.2$ (red marker) and seeks to cancel the waves by creating destructive interference. The controller is informed by three sensors (orange markers), and its performances are measured in the interval $[x^*_1,x^*_2]$ in terms of $l_2$ norm of the solution. The figure shows three snapshots at three episodes $\mbox{ep}=2,5,15$. In a dozen episodes, the control is excellent.}\label{Fig6}
\end{figure}

The architecture implemented in this example is shown in Figure \ref{Fig5}, on the right. The policy network is connected to the Q network. The environment to be controlled is a 1D linear advection equation subject to a disturbance and a control action:

\begin{equation}
\label{eq53}
\frac{\partial u}{\partial t} + 330\frac{\partial u}{\partial x} = {\color{red}\underbrace{ 300 \sin{(100\pi t)}\cdot \mathcal{N}(x-5,0.2)}_{disturbance}} +  {\color{blue}\underbrace{ a_t \cdot \mathcal{N}(x-18.2,0.2)}_{control}}\,,
\end{equation} over a domain $x \in [0, 50]$. Both sides of the domain are open (with non-reflecting conditions). The domain and a snapshot of the system is illustrated in Figure \ref{Fig6}. The disturbance term consists of a Gaussian function multiplied by a sinusoid with given frequency and amplitude. The control term consist of an identical Gaussian placed downstream and having amplitude driven by the agent (controller).


The amplitude $a_t$ constitutes the action of the agent. These are taken as a function of the system state $s_t$, which in this case consists of a vector of $3$ points sampled at three locations. The agent's goal is to cancel the disturbance downstream of its location, i.e. achieving $u(x>18.2)\approx 0$. The reward is defined as $r_t=-||u_O(t,x)||_2$, with $u_O$ the sampled solution in the interval $x^{*}_1$-$x^{*}_2$, also indicated in Figure \ref{Fig6}.

The architecture of the two ANNs is also detailed in Figure \ref{Fig5}, on the right. The policy is an ANN with three inputs, two hidden layers of 128 neurons and one output. Relu activation functions are used in all layers. The Q network has 4 inputs and two joining layers that merge into two hidden layers of 128 neurons each. More details on the implemented structures are available in \cite{Pino}.

Figure \ref{Fig6} shows three snapshots of the solution at three episodes, namely $\mathbf{ep}=2,5,15$. The control performance is excellent within dozens of episodes, and the incoming waves are nearly entirely cancelled. While it is easy to show that a linear controller could easily solve this problem, no information about the physics of the system has been used in the RL approach.

\section{Conclusions and Perspectives}
\label{sec:Sec6}

This talk explored avenues for machine learning in fluid mechanics. Machine learning methods are designed to learn functions from data without leveraging on first principles. The fundamentals of supervised, unsupervised and reinforcement learning were briefly reviewed, and several representative algorithms were tested on problems of fluid mechanics.

Within the supervised learning category, deep learning and genetic programming were shown to outperform classic tools in tasks such as thermal turbulence modelling and noise prediction. These tools can be seen as adaptive function approximators, and care is needed in introducing physical priors in their calibration. This can be done upfront, when constructing the learning architecture, or during the learning, incorporating them in the cost function and its constraints. This is a natural approach considering that most machine learning models (ANN, RBFs, GP trees) are easily differentiable and analytic constraints in the form of PDEs can be easily included in the learning process. Given their predictive capabilities and the availability of powerful open-source libraries for their implementation, it is believed that ANN will become essential tools in the toolbox of the fluid dynamicist. 

Besides allowing for ``informing'' the regression about physical priors, the differentiability of machine learning models also enables the meshless integration of PDEs. An example using RBFs to compute pressure fields from LPT measurements was illustrated. These methods allow avoiding the difficulties of meshing and numerical integration in the presence of noise, limited resolution and complex geometries, as in the case in the pressure integration from image velocimetry. In the authors' opinion, these tools will play a prominent role in data assimilation and experimental fluid mechanics.

Within the unsupervised learning category, linear and nonlinear tools for dimensionality reduction were showcased. While the fluid mechanic's community has vast experience on linear methods (and has pioneered some of the most powerful decompositions), the use of nonlinear methods is in its infancy. The provided illustration showed the vastly superior `data compression' performances of the kPCA over the PCA on a TR-PIV dataset. In tasks that do not demand physical interpretation, such as data compression or filtering, nonlinear tools and manifold learning techniques will likely become the standard tools. These are also precious pre-processing tools for classification and time series forecasting. On the other hand, the performances of nonlinear tools are very sensitive to hyper-parameters (e.g. the kernel parameters), and the identified manifolds might be hard to interpret.

Finally, the bridge between optimal control theory and reinforcement learning (RL) was highlighted. RL algorithms are designed to solve decision problems via trial and error and have proved capable of outstanding achievements in board games. In the provided example and the recent literature, successful applications of RL for flow control have been illustrated. These entitle the community to hope that a RL agent might one day learn control strategies in the same way it can nowadays learn how to play chess at superhuman levels. In such a scenario, which today appears on the edge of science fiction for fluid mechanics, the usual role might be inverted, and the computer (agent) would become our supervisor. This is already happening in board games. But the problems of controlling turbulence is vastly more complex than any board game, and RL appears somewhat a less mature field than supervised or unsupervised learning. New training agents are published every year and, at the time of writing, none of the most popular algorithms (DDPG, PPO or A3C) is older than six years. The field is open and is looking for a new generation of engineers trained in both fluid mechanics \emph{and} machine learning.

\bibliographystyle{apalike}
\bibliography{Mendez_et_al}
\clearpage{\pagestyle{empty}\cleardoublepage}
\end{document}